\documentclass[usenatbib]{mn2e}
\usepackage{amssymb}
\usepackage{amsmath}
\usepackage{array}
\usepackage{bigstrut}
\usepackage{cellspace}
\usepackage{graphicx}

\newif\ifAMStwofonts
\def\ut #1 #2 { \, \rmn{#1}^{#2}}
\def\grad{\bmath{\nabla}}
\def\cross{\bmath{\times}}
\def\bdot{\bmath{\cdot}}
\def\curl #1 {\grad \cross #1}
\def\div #1 {\grad \bdot #1}

\def\v{\bmath{v}}

\def\B{\bmath{B}}
\def\E{\bmath{E}}            

\def\zh{\bmath{\hat{z}}}     

\def\vr{v_{r}}               
\def\vk{v_{\rm K}}           

\def\J{\bmath{J}}

\newcommand{\ee}[1]{\times 10^{#1}}

\usepackage[]{graphics}
\title[Magnetic polarity in extragalactic jet sources]{On the
interpretation of the apparent existence of a preferred magnetic
polarity in extragalactic jet sources}
\author[A. K\"onigl]
{Arieh K\"onigl$^{1}$\thanks{E-Mail: akonigl@uchicago.edu}\\
$^1$Department of Astronomy \& Astrophysics and The Enrico Fermi
Institute, The University of Chicago, Chicago, IL 60637, USA}

\pagerange{\pageref{firstpage}--\pageref{lastpage}}
\pubyear{2010}
\begin{document}

\maketitle
\label{firstpage}
\begin{abstract}

Contopoulos et al. recently argued that there is observational evidence
for a preferred sense of the Faraday rotation-measure gradients across
jets from active galactic nuclei (AGNs). Such behaviour could arise if
there were a deterministic relationship between the polarity of the
poloidal magnetic field that threads the outflow and the sense of
rotation of the outflow's source.  Based on this interpretation,
Countopoulos et al. suggested that their finding supports a model for
the origin of cosmic magnetic fields in a Poynting-Robertson process
operating in AGN accretion discs. Here I point out that an alternative
explanation of such a relationship could be that the Hall current plays
a key role in the magnetohydrodynamics of the underlying disc. In this
picture, the measured Faraday rotation is dominated by the contribution
of a centrifugally driven wind that is launched from the weakly ionized
outer region of the disc. Additional observations are, however, needed
to verify the claimed behaviour.  

\end{abstract} 


\section{Introduction}
\label{sec:intro}

In a recent paper, \citet[][hereafter CCKG09]{Con09} compiled data on
transverse gradients in the Faraday rotation measure (FRM) across jets
associated with active galactic nuclei (AGNs). The data were obtained by
high-resolution (milliarcseconds, typically corresponding to a projected
scale of parsecs at the source) multi-wavelength radio polarization
observations. They found that the majority (22/29) of sources in which
gradients were detected relatively close to the base of the jet
exhibited clockwise gradients. As explained by CCKG09, such gradients
are expected in outflows that contain a helical magnetic field (see
Fig.~1 in \citealt{Gab08}). A helical field arises naturally when
poloidal field lines that thread the outflow are twisted by the
differential rotation of the source (an accretion disc or the central
black hole in the case of AGNs), and the sense of the twist (as
reflected in the FRM gradient) is directly related to the relative
orientation of the poloidal field and the rotation vector (see Fig.~2 in
\citealt{GVMO08}). The inferred ordered magnetic field could originate
in stars or interstellar gas and be dragged into the jet launching
region by the accretion flow, or it could be generated in the disc by a
dynamo mechanism. In either case, one would expect an equal probability
for the poloidal field component to have positive or negative
polarity. (For definiteness, I take the source to be a disc whose
rotation vector points along the $+\zh$ direction of a cylindrical
coordinate system $\{r, \phi, z \}$ placed at the disc's centre, with
the polarity of the magnetic field $\B$ defined as ${\rm sgn}\{B_z\}$.)
CCKG09 suggested that the low ($< 1\%$) probability for the compiled
distribution to be a chance occurrence indicates that there is a
physical mechanism that acts to impose a {\em positive} magnetic
polarity (i.e. $B_z$ parallel to the rotation vector) at the source of
AGN jets.

CCKG09 proposed that the relevant mechanism is the Poynting-Robertson
`Cosmic Battery' effect \citep{CK98}. In this picture, the radiation
drag induced by the central AGN radiation field and acting predominantly
on the electrons in a rotating circumnuclear disc gives rise to an
azimuthal current that generates a poloidal magnetic field with a unique
(positive) polarity. Although the viability of this mechanism has been
debated in the literature \citep[e.g.][]{BKB77,BKLB02,CKC06,CCK08},
CCKG09 argued that the indicated preponderance of positive polarities in
AGN jets could be an important factor in its favour in view of the
apparent difficulty of explaining this behaviour `using any standard MHD
model in the literature'. Here I point out that, in fact, a single-
(positive) polarity disc outflow can naturally form also under `standard
MHD' conditions if the Hall current affects the magnetohydrodynamics
(MHD) of the associated accretion flow. In this picture, the gas that
dominates the observed FRMs corresponds to a wind that originates in the
comparatively massive and weakly ionized outer region of the disc. If
the claimed effect is real (which, in view of the considerable
observational difficulties involved, requires further confirmation) and
the disc-wind interpretation is correct then the FRM measurements could
yield valuable clues to the physical conditions in the outflow as well
as in the underlying disc.

\section{Hall-Current Effects in Wind-Driving Discs}
\label{sec:Hall}

For the purpose of illustration, I consider a simplified model of a
wind-driving Newtonian disc that is in nearly Keplerian rotation (with
azimuthal speed $|v_\phi| \approx \vk$) around a black hole of mass $M$,
vertically isothermal (with sound speed $c_{\rm s}$) and geometrically
thin (so that the density scaleheight $h(r) \ll r$). I focus on systems
that are axisymmetric and in a steady state on the dynamical time
$\Omega_{\rm K}^{-1}(r)= r/v_{\rm K}(r)$. I consider outflows that are
launched in the form of a centrifugally driven wind
\citep[e.g.][]{BP82}, which are also relevant to jets that attain
relativistic speeds \citep[e.g.][]{VK04}. The disc is assumed to be
threaded by a large-scale, open magnetic field with an `even' $B_z$
symmetry about the mid-plane (corresponding to $B_r=B_\phi=0$ at
$z=0$). On going away from the mid-plane, the poloidal field lines bend
away from the rotation axis, and the resulting radial component ($B_r$)
is sheared by the disc's differential rotation to produce an azimuthal
component ($B_\phi$). The radial and azimuthal field components
generated in this way satisfy $B_r B_\phi < 0$. The $B_\phi$ component
induces a torque ($\propto r B_z B_\phi$) that acts to brake the disc
rotation. The magnetic field removes angular momentum from the disc,
and, if the field-line inclination at the disc's surface is large enough
($B_r/B_z > 1/\sqrt{3}$), a wind can be launched wherein the field
transfers the angular momentum back to the gas (thereby accelerating it
`centrifugally').

The azimuthal shearing as well as the tendency of the accretion flow to
advect the magnetic field lines inward are countered by the magnetic
diffusivity of the gas. However, the diffusivity cannot be too large if
vertical magnetic transport of angular momentum is to take place; in
fact, the minimum-coupling condition is essentially the same as the
corresponding condition for efficient radial magnetic transport through
turbulence induced by the magnetorotational instability \citep[MRI;
see][]{KS11}. In the region where this condition is satisfied, MRI may
dominate at low disc elevations, with vertical transport by the
large-scale field taking over further up \citep[][hereafter
SKW07]{SKW07}. However, to simplify the discussion, I assume that the
degree of coupling and the magnetic field strength are high enough for
vertical transport to dominate already at the mid-plane. Wind-driving
disc models of this type were previously studied in the context of
protostellar systems (e.g. \citealt{WK93}; \citealt{KSW10};
\citealt{SKW10}; hereafter WK93, KSW10 and SKW10, respectively).

\begin{figure*}
\centering
\includegraphics[height=6.2cm]{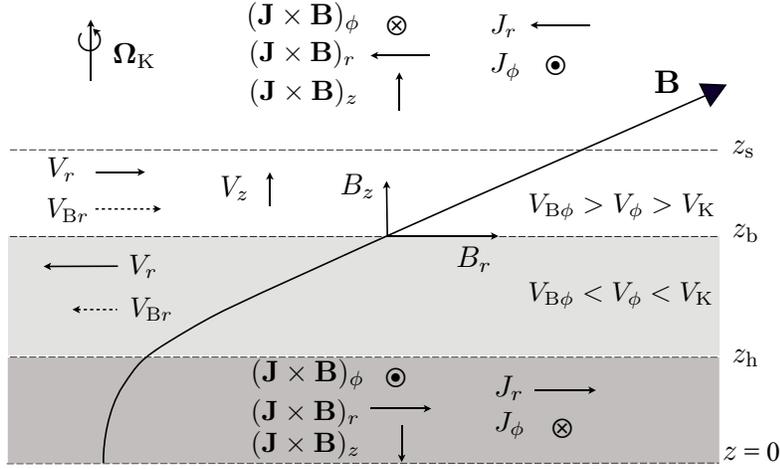}
\caption{Schematic of the vertical structure of a wind-driving disc,
showing a representative field line and the orientations of the
Keplerian angular velocity $\mathbf{\Omega}_{\rm K}$ and of components
of the gas velocity ($\mathbf{v}$), field-line velocity
($\mathbf{v}_{\rm B}$), magnetic field ($\mathbf{B}$), current density
($\mathbf{J} \propto \mathbf{\nabla \times B}$) and magnetic force
($\propto \mathbf{J\times B}$) vectors in the disc ($z< z_{\rm b}$) and
wind ($z>z_{\rm b}$) regions. See text for further details.}
\label{fig1}
\end{figure*}

The vertical structure of the envisioned wind-driving disc region is
shown schematically in Fig.~\ref{fig1}. One can identify three basic
zones (WK93): the quasi-hydrostatic layer near the mid-plane (between
$z=0$ and $z_{\rm h} = h$) where the bulk of the matter is concentrated
and most of the field-line bending takes place, a transition zone where
the magnetic pressure comes to dominate the thermal pressure and the
inflow gradually diminishes, and an outflow region (between the top of
the disc at $z_{\rm b}$ and the `sonic' critical surface at $z_{\rm s}$)
that corresponds to the base of the wind (with the mass outflow rate
effectively determined by imposing the regularity condition at $z_{\rm
s}$). The magnetic field lines are tied (possibly imperfectly) to the
ionized disc component, and the collisional drag between this component
and the neutrals transmits the magnetic torque to the bulk of the
gas. Therefore $v_{\rm B\phi} < v_\phi$ inside the disc. The loss of
angular momentum enables the neutrals to drift toward the centre,
resulting in a radial drag on the ionized component that tends to impart
an inward radial speed $|v_{{\rm B}r}|<|v_r|$ to the field lines. The
radial drag is balanced by the magnetic tension exerted by the
outward-bent field lines, and this force, in turn, contributes to the
radial support of the neutral gas inside the disc and causes it to
rotate at sub-Keplerian speeds ($v_\phi < \vk$). Outside the disc the
situation is reversed: the field lines transfer angular momentum to the
matter ($v_{\rm B \phi} > v_\phi$), which consequently rotates at
super-Keplerian speeds ($v_\phi > \vk$).

The effect of the magnetic tension on the disc gas can be seen
explicitly from the $r$ component of the momentum equation, which in the
quasi-hydrostatic region of a thin disc (in which the $v_z$ velocity
component can be neglected and vertical gradients almost always dominate
radial gradients) can be approximated by 
\begin{equation}
\frac{2 \rho \vk}{r} (\vk - v_{\phi}) \approx \frac{J_\phi B_z}{c}
\label{eq:rmom} 
\end{equation} 
(KSW10), where $\rho$ is the mass density, $c$ is the speed of light and
the term involving the current density $J_\phi \approx (c/4\pi)\partial
B_r/\partial z$ represents the magnetic tension force. Under the same
approximations, and assuming that angular momentum transport by the
large-scale field dominates, the $\phi$ component of the momentum
equation can be written as
\begin{equation}
\frac{\rho \vr \vk}{2r} \approx - \frac{J_r B_z}{c}\ ,
\label{eq:fmom}
\end{equation}
where $J_r \approx - (c/4\pi)\partial B_\phi/\partial z$. Combining
equations~(\ref{eq:rmom}) and~(\ref{eq:fmom}) yields
\begin{equation}
\left ( \frac{d B_r}{d B_\phi} \right )_0 = \frac{4(\vk-v_{\phi 0})}{v_{r0}}<0
\label{eq:dbrdbf1}
\end{equation}
(where the subscript `0' denotes the mid-plane), which corroborates the
inference that $B_rB_\phi<0$ in the wind-driving region of the disc.

The effect of the magnetic diffusivity on the disc structure can be
inferred from a consideration of Ohm's law, which for a weakly ionized
gas takes the form
\begin{equation}
\E = - \frac{\v \cross \B}{c} + \frac{4\pi}{c^2}\left [\eta_{\rm O}
\J + \eta_{\rm H} \frac{\J\cross\B}{B} - \eta_{\rm A}
\frac{(\J\cross\B)\cross\B}{B^2}\right ]\, ,
\label{eq:ohm}
\end{equation}
where $\eta_{\rm O}$, $\eta_{\rm H}$ and $\eta_{\rm A}$ are,
respectively, the Ohm, Hall, and ambipolar diffusivities. The latter can
be evaluated for an arbitrary charge composition, but it is often a good
approximation to assume that the current carriers consist only of
electrons (subscript `e') and singly charged ions (subscript `i') with a
mass ratio $m_{\rm e}/m_{\rm i} \ll 1$. In this case one can write
$\eta_{\rm O} = c^2/4\pi \sigma_{\rm e}$, $\eta_{\rm H} = \beta_{\rm e}
\eta_{\rm O}$ and $\eta_{\rm A} = \beta_{\rm e} \beta_{\rm i} \eta_{\rm
O}$ \citep{War07}, where $\sigma_{\rm e} = {n_{\rm e} e^2}/{m_{\rm e}
\nu_{\rm e}}$ is the electron conductivity and
\begin{equation}
\beta_{\rm j}\equiv \frac{eB}{m_{\rm j} c} \, \frac{1}{\nu_{j}}
\label{eq:Hall_parameter}
\end{equation}
is the Hall parameter of a charged species j. In these expressions,
$n_{\rm e}$ is the electron number density (which is equal to $n_{\rm
i}$), $e$ is the electronic charge, $\nu_{\rm j}=\rho <\sigma v>_{\rm
jn}/(m_{\rm j} + m_{\rm n})$ is the momentum-exchange collision
frequency with the neutral (subscript `n') particles (where $<\sigma
v>_{\rm j n}$ is the charge--neutral momentum transfer rate) and $B
\equiv |\mathbf{B}|\, {\rm sgn}\{B_z\}$ is the {\em signed} magnetic
field amplitude. The absolute value of $\beta_j$ represents the ratio of
the cyclotron frequency $\omega_{{\rm c} j}$ to $\nu_j$ and is a measure
of whether the corresponding charged particle is well ($|\beta_j|\gg 1$)
or poorly ($|\beta_j|\ll 1$) coupled to the magnetic field. Using
$m_{\rm i} \approx 30 \, m_{\rm H}$ \citep*[e.g.][]{DRD83}, the ratio of
the ion and electron Hall parameters can be estimated to be $q \equiv
\beta_{\rm i}/\beta_{\rm e} \approx 1.3\ee{-4} \sqrt{T}$ (where $T$ is
the gas temperature), which is typically $\ll 1$.

Note that, among the three diffusivity terms, only $\eta_{\rm H}$
depends on an odd power of $B$ -- this is the origin of the dependence of
the disc solutions on the magnetic field polarity when the Hall current
is dynamically important. To demonstrate the fact that negative-polarity
field configurations are excluded in certain parameter
regimes, consider the $\phi$ component of equation~(\ref{eq:ohm}) in the
quasi-hydrostatic zone. As discussed in KSW10, $E_\phi \approx
E_{\phi0}= v_{{\rm B}r0}B_z/c$ can be set equal to zero in this analysis
without loss of generality (see their Appendix~A, where it is also
argued that physical consistency requires one to consider all the terms
in Ohm's law even in the nominally Hall-dominated regime). The advective
term is approximately $v_rB_z/c$, where $v_r$ can be expressed in terms
of $J_r$ using equation~(\ref{eq:fmom}). Evaluating at $z=0$, one gets
\begin{equation}
\left ( \frac{d B_r}{d B_\phi} \right )_0 = -\frac{J_{\phi 0}}{J_{r 0}}
= - \frac{2a_0^2 + \tilde \eta_{\rm H 0}}{\tilde \eta_{\rm O0}+\tilde
\eta_{\rm A0}} = - \frac{2\Upsilon_0 + \beta_{\rm i
0}^{-1}}{1+\beta_{\rm e0}^{-1}\beta_{\rm i0}^{-1}}\ ,
\label{eq:dbrdbf2}
\end{equation}
where the third expression is general and the fourth one applies in the
case of a two-component plasma with $q\ll 1$. In equation~(\ref{eq:dbrdbf2}),
the diffusivities are normalized using $\tilde \eta
\equiv \eta/h_{\rm T} c_{\rm s}$, where $h_{\rm T} = c_{\rm
s}^2/\Omega_{\rm K}$ is the tidal scaleheight, and the parameters
\begin{equation} 
a_0 \equiv \frac{v_{{\rm A}0}}{c_{\rm s}}\ ,
\label{eq:a_0}
\end{equation}
the mid-plane ratio of the Alfv\'en speed $v_{\rm A} = |\mathbf{B}|/\sqrt
{4 \pi \rho}$ to the isothermal sound speed, and
\begin{equation}
\Upsilon_0 \equiv \frac{\nu_{\rm i 0}}{\Omega_{\rm K}} \frac{m_{\rm
i}n_{\rm i 0}}{\rho_0}\ ,
\label{eq:upsilon}
\end{equation}
the mid-plane ratio of the Keplerian rotation time to the neutral--ion
momentum exchange time, are generally $\lesssim 1$ and $\gtrsim 1$,
respectively (see KSW10). The two terms in the numerator of the third
(or fourth) expression in equation~(\ref{eq:dbrdbf2}) arise,
respectively, from advective and Hall terms in Ohm's law. Their ratio is
given by
\begin{equation}
\frac{\rm Hall}{\rm advective} = \frac{|\tilde{\eta}_{\rm H 0}|}{2a_0^2}=
\frac{1}{2\Upsilon_0|\beta_{\rm i 0}|} = \frac{\Omega_{\rm
K}}{2\omega_{\rm H 0}}\ ,
\label{eq:ratio}
\end{equation}
where the last expression involves the Hall frequency
\begin{equation}
\omega_{\rm H} \equiv \frac{n_{\rm i}}{\mu n}\, \omega_{\rm c p}\, ,
\label{eq:omega_H}
\end{equation}
whose inverse represents the characteristic time-scale of the Hall
effect in MHD \citep[e.g.][]{PW08}. (Here $n$ is the particle number
density, $\mu$ the molecular weight and the subscript p denotes a
proton.) When the ratio~(\ref{eq:ratio}) is $>1$ and ${\rm
sgn}\{B_z\}<0$, equation~(\ref{eq:dbrdbf2}) implies that $({d B_r}/{d
B_\phi})_0$ is $>0$, which is inconsistent with the inference from the
physical model (equation~\ref{eq:dbrdbf1}). Therefore, when this ratio
is $>1$, {\em only positive-polarity disc outflows can form}.

Specializing to the case of a two-component plasma with $q\ll 1$, the
Hall-dominance condition can be written as
\begin{equation}
\frac{\Omega_{\rm K}}{2\omega_{\rm H 0}} = \frac{\mu_{\rm
d}}{2}\frac{\Omega_{\rm K}}{\omega_{\rm c p 0}}\frac{1}{x_{\rm i d 0}}
= \frac{8.2 \times 10^{-13}}{x_{\rm i d 0}} 
\frac{M_8^{1/2}}{B_{0.1}r_{0.1}^{3/2}} > 1\, ,
\label{eq:ratio1}
\end{equation}
where the subscript `d' denotes the disc, $x_{\rm i d}\equiv n_{\rm i
d}/n_{\rm d}$ is the degree of ionization of the disc gas, $\mu_{\rm d}
\approx 2.33$, $M_8 \equiv M/{10^8\, \rm M}_\odot$, $B_{0.1}\equiv
B_0/0.1\, {\rm G}$ and $r_{0.1} \equiv r/0.1\, {\rm pc}$. When the
parameter constraints on physically viable wind-driving molecular discs
are analyzed in detail (see WK93, KSW10 and SKW10), one finds that the
condition~(\ref{eq:ratio1}) is satisfied in two (out of the four
identified) parameter regimes in the nominal Hall diffusivity domain
($|\beta_{\rm i}| \ll 1 \ll |\beta_{\rm e}|$) and in one regime (out of
the three identified) in the nominal Ohm domain ($|\beta_{\rm e}| \ll
1$). These are precisely the parameter regimes that admit only
positive-polarity disc/wind solutions. (In the other viable-solution
regimes both polarities are allowed.) In what follows, I consider
whether the inequality~(\ref{eq:ratio1}) can be satisfied in AGN discs
but do not address the additional parameter constraints derived in the
above-cited references, which could, however, be useful in a more
detailed treatment.

\section{Application to AGN Discs}
\label{sec:apply}

Although the physical variables that appear in equation~(\ref{eq:ratio1})
cannot be directly determined from existing observations, one can
nevertheless extract useful clues from the Faraday rotation measurements
themselves. For example, \citet{Kha09} deduced from observations of
parsec-scale jets in Fanaroff-Riley Type I (FR I) radio galaxies that
the Faraday rotation is likely associated with a helical (or toroidal)
magnetic field and that it arises in a `sheath' around the jet, which
could correspond to a magnetized disc wind \citep[e.g.][]{BL95}. They
inferred the magnetic field amplitude by equating it to the
equipartition field ($\sim 5\, {\rm mG}$) derived from the synchrotron
radio emission in the jet, which is reasonable if the jet is confined by
the disc wind and both represent MHD outflows that are magnetic pressure
dominated. Using a representative FRM value of $200\ {\rm rad}\ {\rm
m}^{-2}$ and a pathlength of $\sim 1\ {\rm pc}$ through the sheath, they
estimated a characteristic electron density of $\sim 0.03\; {\rm
cm}^{-3}$. This value could plausibly be ascribed to the atomic hydrogen
layer that arises in the innermost zone of a photodissociation region
formed when the AGN ionizing radiation (assumed not to be dominated by hard
X-rays) impinges on the disc outflow \citep[e.g.][]{TH85}. A layer of
this type is established just beyond the HII/HI transition zone (whose
column density is $\lesssim 10^{20}\, {\rm cm}^{-2}$) and has a
characteristic electron fraction of $\sim 3 \times 10^{-4}$ (resulting
primarily from singly ionized carbon atoms). This, in turn, implies a
density $n_{\rm w} \approx 10^2\, {\rm cm}^{-3}$ (where the subscript
`w' denotes the wind) within the sheath.

Now, if the parsec-scale sheath with $|B| \lesssim 10\, {\rm mG}$ can be
identified with a centrifugally driven disc wind, the outflow likely
originates on scales $r \lesssim 0.1\, {\rm pc}$, where the magnetic
field strength is $B(r) \lesssim 0.1\, {\rm G}$, and its speed $v_{\rm
w}$ is $\gtrsim v_{\rm K}(r)$ \citep[e.g.][]{BP82}. I adopt $r_{0.1} =
1$ and $B_{0.1}=1$ as fiducial values and use $M_8 = 3$ as a
representative mass for radio-loud AGNs \citep[e.g.][]{SSL07}. The mass
outflow rate in the sheath can be estimated to be $\sim r_{\rm w}^2
\mu_{\rm w} m_{\rm p} n_{\rm w} v_{\rm w} \approx 8\times 10^{23}\, {\rm
g}\, {\rm s}^{-1}$ (using $r_{\rm w} \approx 1\, {\rm pc}$, $v_{\rm w}
\approx v_{\rm K}(r_{0.1}=1)$ and $\mu_{\rm w} \approx 1.27$). Given
that an efficient disc wind typically has a mass outflow rate that is
about an order of magnitude smaller than the accretion rate
\citep[e.g.][]{KS11}, I adopt $\dot M_{25}\equiv \dot M /10^{25}\, {\rm
g}\, {\rm s}^{-1} \approx 1$ (corresponding to $\dot M \approx 0.16\, {\rm
M}_\odot\, {\rm yr}^{-1}$) for the mass accretion rate at
$r_{0.1}=1$. The mid-plane-to-disc-surface column density is given by $N
= \dot M/4\pi r \mu_{\rm d} m_{\rm p} |v_{r 0}|$, and, by setting $h
\approx h_{\rm T}$ (an upper limit due to the magnetic squeezing of the
disc; see KSW10 and Fig~\ref{fig1}), a lower limit on the mid-plane
density can be obtained by writing $n_{\rm d 0} = (2/\pi)^{1/2}N/h_{\rm
T}$. Using $|v_{r 0}| = 0.1\, \epsilon_{0.1}\, c_{\rm s}$ (where
$\epsilon_{0.1}$ is typically $\gtrsim 1$; e.g. SKW07) and adopting
$T_{500}\equiv T_0/500\, {\rm K} \approx 1$ as a rough estimate of the
mid-plane temperature at the radius of interest, I get $N \approx
5\times 10^{25}\, \dot M_{25}\, \epsilon_{0.1}^{-1}\, r_{0.1}^{-1}\,
T_{500}^{-1/2}\ {\rm cm}^{-2}$ and $n_{\rm d 0} \gtrsim 3\times
10^{11}\, (M_8/3)^{1/2}\, \dot M_{25}\, \epsilon_{0.1}^{-1}\,
r_{0.1}^{-5/2}\, T_{500}^{-1}\ {\rm cm}^{-3}$.

For the adopted values of $M_8$, $r_{0.1}$ and $B_{0.1}$, the condition
given by equation~(\ref{eq:ratio1}) can be fulfilled if $x_{\rm i d
0}\lesssim 10^{-12}$. The dominant ionization agent for the inferred
values of $r$ and $N$ would likely be cosmic rays, and if one employs
the nominal Galactic values for the ionization rate ($\xi_{\rm CR}
\approx 1\times 10^{-17}\, {\rm s}^{-1}$), the cosmic-ray attenuation
grammage ($\chi \approx 100\, {\rm g\, cm}^{-2}$) and the gas-to-dust
mass ratio ($\approx 100$), one finds that the requirement on $x_{\rm i
d 0}$ should be satisfied for the modelled disc
\citep[e.g.][]{UN90}. Although the charged-particle abundances depend on
the uncertain heavy-element depletions and grain size distribution, ions
and electrons have a significant influence on the magnetic diffusivity
under typical Galactic conditions so long as $(n/\xi_{\rm
CR})\exp{(\mu_{\rm d} m_{\rm p} N/\chi)}$ is $\lesssim 10^{29}\, {\rm
cm}^{-3}\,$s \citep[e.g.][]{KM10}. At higher values of this parameter
combination, singly charged (+ and -) grains of equal mass become the
dominant current carriers, which results in the Hall conductivity
tending to zero (so Hall effects no longer play a significant role). The
actual value of $\xi_{\rm CR}$ in AGN discs is still unknown: on the one
hand it might be substantially higher than in the Galaxy
\citep[e.g.][]{Bay09}, but on the other hand a super-Alfv\'enic disc
outflow and the expected magnetic field geometry of the disc/wind system
could strongly reduce the cosmic-ray flux reaching the mid-plane
\citep[e.g.][]{War07}. In view of these uncertainties, the estimates
employed in the evaluation of the inequality~(\ref{eq:ratio1}) need to
be regarded as merely illustrative. These estimates nevertheless indicate
that the proposed scenario is not implausible. Note that the
wind-launching region is likely to contain magnetic fields of either
polarity, both in the case in which the field is advected from a larger
distance (for example, if it originates in stars with randomly oriented
magnetic moments whose remnants are brought into the launching region in
sufficiently rapid succession) and in the case of a local disc dynamo
\citep[e.g.][]{BB03}. Therefore, when the condition~(\ref{eq:ratio1}) is
satisfied, a positive-polarity outflow should in general be able to
form.

\section{Discussion}
\label{sec:discuss}

In the interpretation outlined in this Letter, the Faraday sheath that
was inferred to be the likely source of the measured parsec-scale FRM
gradients is identified with a centrifugally driven wind that originates
in a weakly ionized, molecular region of an AGN accretion disc at a
distance $\gtrsim 0.1\;$pc from the central black hole. This picture is
consistent with the idea that an MHD disc outflow constitutes the `dusty
torus' that underlies AGN unification schemes \citep[][]{KK94}. The
suggestion that the gas in the outer regions of certain AGN discs is
mostly molecular and dusty is supported by several observational
findings. Although molecular line measurements do not yet resolve these
regions, there are already strong indications that dense, rotating
molecular gas is concentrated on scales $\gtrsim 10\;$pc around the
centres of local AGNs \citep[e.g.][]{Hsi08,Hic09}, and numerical
simulations \citep[e.g.][]{WPS09,Sch10} indicate that this gas likely
forms a thin disc on smaller scales. Flattened distributions of dust
have in fact been imaged in sources like NGC 1068 on $\gtrsim 1\;$pc
scales by infrared observations \citep[e.g.][]{Rab09}. Direct evidence
for dense ($n \gtrsim 10^8\; {\rm cm}^{-3}$) molecular gas shielded from
the central continuum by a large dusty column ($N \gtrsim 10^{23}\; {\rm
cm}^{-2}$) has been provided in several AGNs (including NGC 1068;
\citealt{GG97}) by the detection of flattened distributions (imaged on
scales $\sim 0.1-1\;$pc) of water masers that exhibit Keplerian (or
nearly Keplerian) rotation-velocity profiles. The maser emission could
arise either in the circumnuclear disc \citep[e.g.][]{NM95} or in dense
clumps within a dusty disc wind \citep[e.g.][]{KKE99}. Although
all the above findings pertain to spiral galaxies (Seyferts
and LINERs), there is also evidence that molecular gas is
present in the centres of nearby radio-loud elliptical galaxies
\citep[e.g.][]{Oca10}. Higher-resolution and more sensitive future
observations, particularly with ALMA, should shed more light on this
issue.

A growing body of data indicates that relativistic AGN jets undergo the
bulk of their acceleration and collimation on spatial scales not much
smaller than $\sim 0.1\;$pc \citep[e.g.][]{Sik05,JBL99}. Disc outflows
could play an important role in these processes because efficient MHD
acceleration to relativistic speeds requires lateral confinement over
extended spatial scales \citep[e.g.][]{Kom09} and because the presence
of a surrounding disc wind could greatly facilitate the collimation of a
relativistic jet, which might otherwise be hard to achieve
\citep[e.g.][]{Gra09}. It can be expected that the portion of the disc
wind that is most effective in confining the jet on scales
$\gtrsim 0.1\;$pc along the flow is launched from the disc surface at a
comparable distance from the centre \citep[see][]{SFS97}. This could
naturally account for the proposed identification of the Faraday-rotation
sheaths of parsec-scale jets with magnetized winds that originate in
the associated AGN discs at $r \gtrsim 0.1\;$pc.

\section*{Acknowledgments}

I am grateful to Eric Blackman, Ioannis Contopoulos, Denise Gabuzda,
Alfred Glassgold, David Hollenbach, Philip Maloney, Eugene Parker,
Christopher Reynolds, Raquel Salmeron, Gregory Taylor, John Wardle,
Nadia Zakamska as well as the reviewer for helpful input. This work was
supported in part by NSF grant AST-0908184 and by the NSF Center for
Magnetic Self-Organization in Laboratory and Astrophysical Plasmas at
the University of Chicago.

\bsp 
\label{lastpage}
\end{document}